\documentclass{desyproc}
\usepackage{epsfig}
\begin{document}
%------------------------------------
\title{Recent soft diffraction results from HERA}

%for single authors the superscripts are optional
\author{{\slshape Alice Valk\'arov\'a\\on behalf of H1 and ZEUS Collaborations}\\[1ex]
Institute of Particle and Nuclear Physics\\
 Faculty of Mathematics 
        and Physics of Charles University\\ 
        V Hole\v sovi\v ck\'ach 2, 180 00 Praha 8,
         Czech Republic}

% if the proceedings are available online (e.g. at Indico)
% please enter the contribution ID or file_name below for the DOI
%\contribID{32}
\contribID{smith\_joe}
\bibliographystyle{unsrt} 
\newcommand\GeV{{\rm ~GeV}}
\def\Pom{{I\!\!P}}
\def\xP{x_{\!\Pom}}
\newcommand{\IP}{I\!\!P}
% TO THE CONFERENCE EDITORS: 
% please update the following information      
% before sending the template to the authors
% \confID{800}  % if the conference is on Indico uncomment this line

\acronym{EDS'09} % if you want the Acronym in the page footer uncomment this line

\maketitle

\begin{abstract}
High statistics measurements of the diffractive reduced cross section $\sigma_r^D$   from the H1 collaboration are presented which make use of two different experimental methods to achieve the largest possible coverage of the kinematic phase space at HERA.  The HERA combined diffractive reduced cross sections based on events with a leading proton are also presented.
%High statistics measurements of the diffractive reduced cross section $\sigma_r^D$ have been made by H1 collaboration using two experimental methods covering accessible kinematic region. The HERA combined diffractive double-differential reduced cross sections in events 
% with a leading  proton were obtained and compared with theoretical predictions.
\end{abstract}

\section{Inclusive H1 measurements}

The study and interpretation of diffraction in $ep$ collisions at HERA provides important inputs for the understanding of quantum chromodynamics (QCD) at high parton densities.
In diffractive interactions the proton stays intact or dissociates into a low mass state ($Y$), while the photon may dissociate into a hadronic state $X$, $\gamma^* p \to Xp' (Y)$.  The systems are separated by a large rapidity gap (LRG).  The diffractive exchange object,$\Pom$, has vacuum quantum numbers and carries a fraction $\xP$ of the initial proton longitudinal momentum and a momentum transfer $t$ from the incoming to the outgoing proton.  
Several theoretical approaches have been proposed to describe the dynamics of diffractive Deep Inelastic Scattering (DIS). A general theoretical framework is provided by the QCD collinear factorisation theorem for DIS cross sections. This implies that the concept of diffractive parton distribution functions (DPDFs) may be introduced. Empirically, an Regge proton vertex factorisation has been found whereby the variables which describe the proton vertex factorise from those describing the hard interaction. The dependencies of $\sigma_r^D$ on $\beta$ and $Q^2$ may then be subjected to a perturbative QCD analysis based on DGLAP equations, in order to obtain DPDFs.

\begin{figure}[htb]
\centerline{%
\includegraphics[width=1.\textwidth]{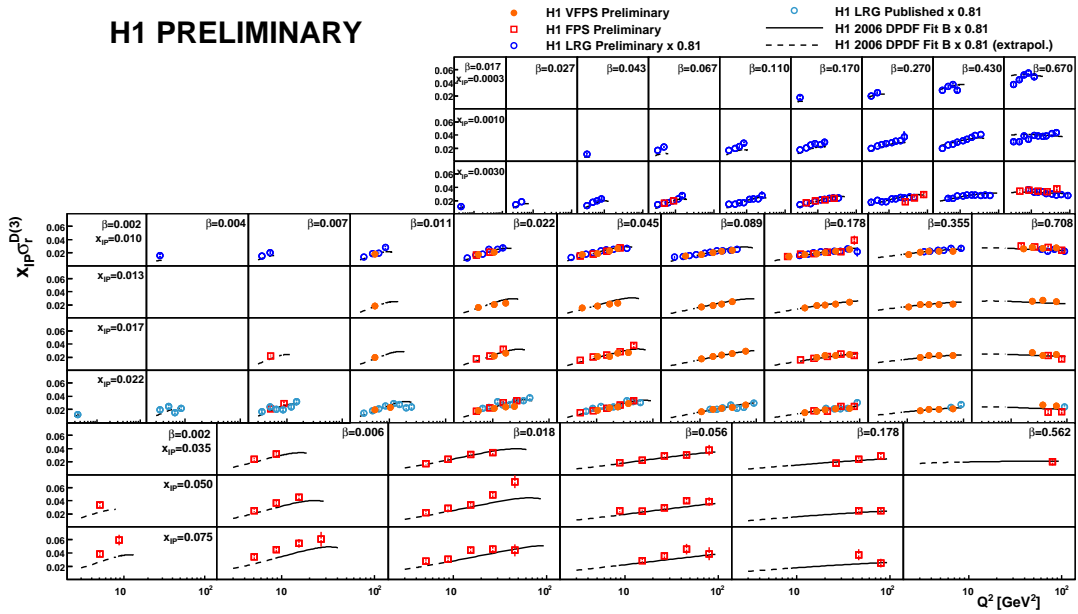}}
\caption{The measurement of diffractive reduced cross section using all experimental methods.}\label{FigureLabel}
\end{figure}

\begin{figure}[htb]
\centerline{%
\includegraphics[width=0.5\textwidth]{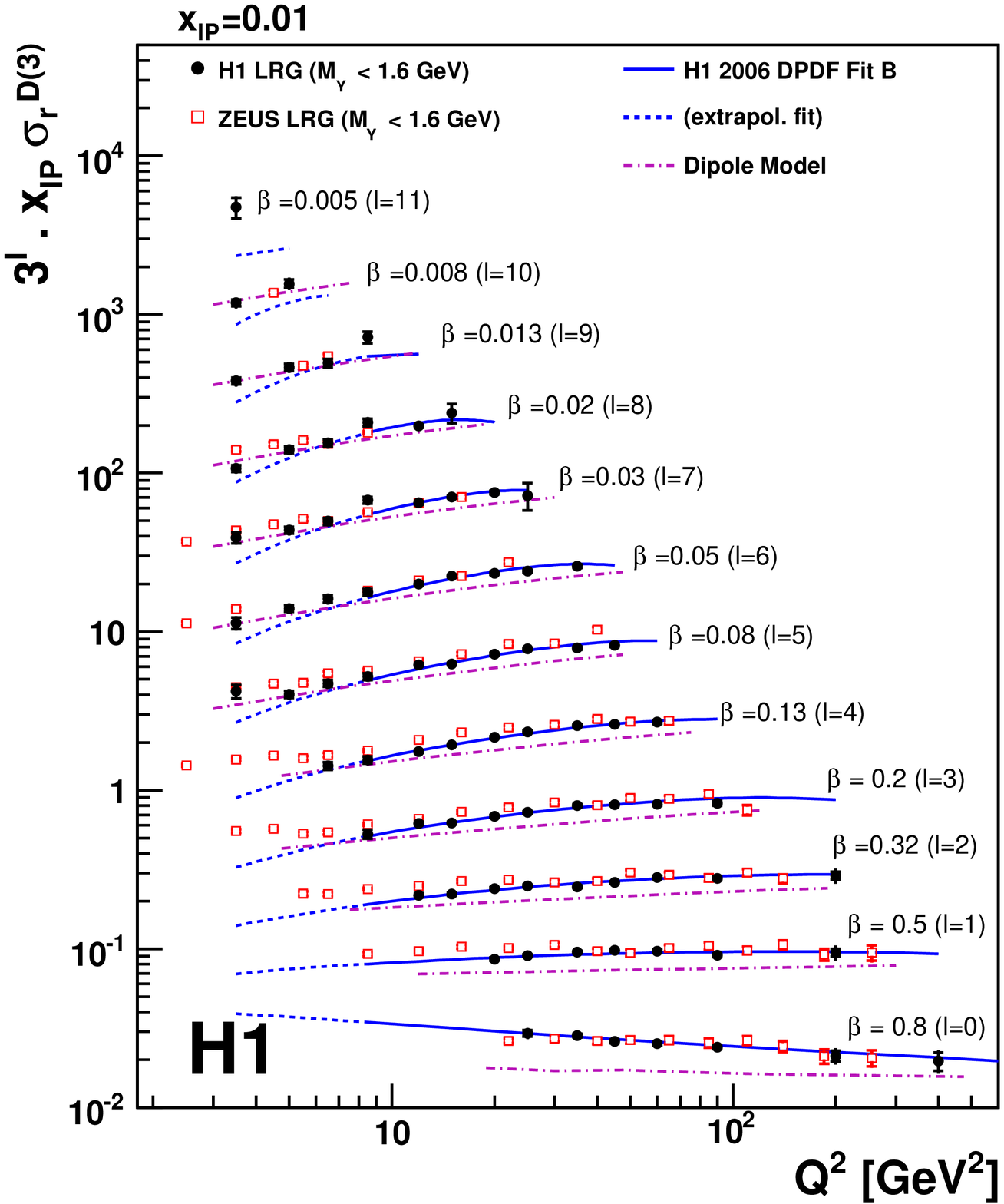}}
\caption{a)~Reduced combined diffractive cross section $\xP\sigma_r^{D(3)}$ compared with results of ZEUS \cite{ZEUS_LRG}, H1 2006 DPDF Fit B predictions and dipole model \cite{dipole}  .}\label{FigureLabel}
\end{figure}

Diffractive events are selected either by detecting the final state proton, or on the basis of a LRG being present. In the latter case, the final system Y escapes detection and the cross section is integrated over ranges in leading baryon mass $M_Y$ and $t$.

\subsection{Diffractive Structure Function Measurements }
Data at the nominal proton beam energy $E_p=920 \GeV$ from the HERA I and II running periods have been analysed to extract diffractive reduced cross section $\sigma_r^D$ in as wide kinematic range as possible, using both LRG and proton detection methods \cite{H1_full}.  In Fig. 1 the full set of measurements as a function of $Q^2$ in bins of $\beta$ is shown. Both the LRG data and DPDF fit B \cite{DPDF} are normalised to $M_Y=M_{proton}$. The data compare well with H1 Fit B prediction. 

A new measurement of the diffractive DIS cross section has been performed using data recorded with the H1 detector in the years 1999-2000 and 2004-2007.  A combination with previous measurements from 1997 based on LRG method  was performed in order to provide a single set of diffractive cross sections for the $Q^2$ range $3 <  Q^2 < 1600$ $\GeV^2$ . The results are shown in Fig.2 and are compared to previously obtained ZEUS cross sections \cite{ZEUS_LRG} as well as to QCD calculations based on DPDFs extracted from H1 data \cite{DPDF} and recent dipole model predictions \cite{dipole}. The predictions of the dipole model including saturation, can describe the low $Q^2$ kinematic domain better than H1 DPDF fits. On the other hand the DPDF fits are more succesful to describe the region of high $Q^2$.
  
\subsection{ Combined HERA measurements}
In Fig. 3 the combined HERA diffractive reduced cross section $\sigma_r^D$ based on the measurements from H1 and ZEUS using the leading proton method is shown
%In Fig.3 is shown 
%the H1 and ZEUS collaborations provided a common analysis of all results obtained for diffractive 
%the result of the common analysis of H1 and ZEUS collaborations, the reduced cross section $\sigma_r^D$ measured using proton spectrometers to detect leading protons 
\cite{H1_ZEUS}. The combination of the measurements results in more precise and kinematically extended diffractive DIS data. The reduction of the total uncertainty of the HERA measurement compared to the input H1 and ZEUS cross sections is clearly visible from Fig. 3. 
  
\begin{figure}[htb]
\centerline{%
\includegraphics[width=0.5\textwidth]{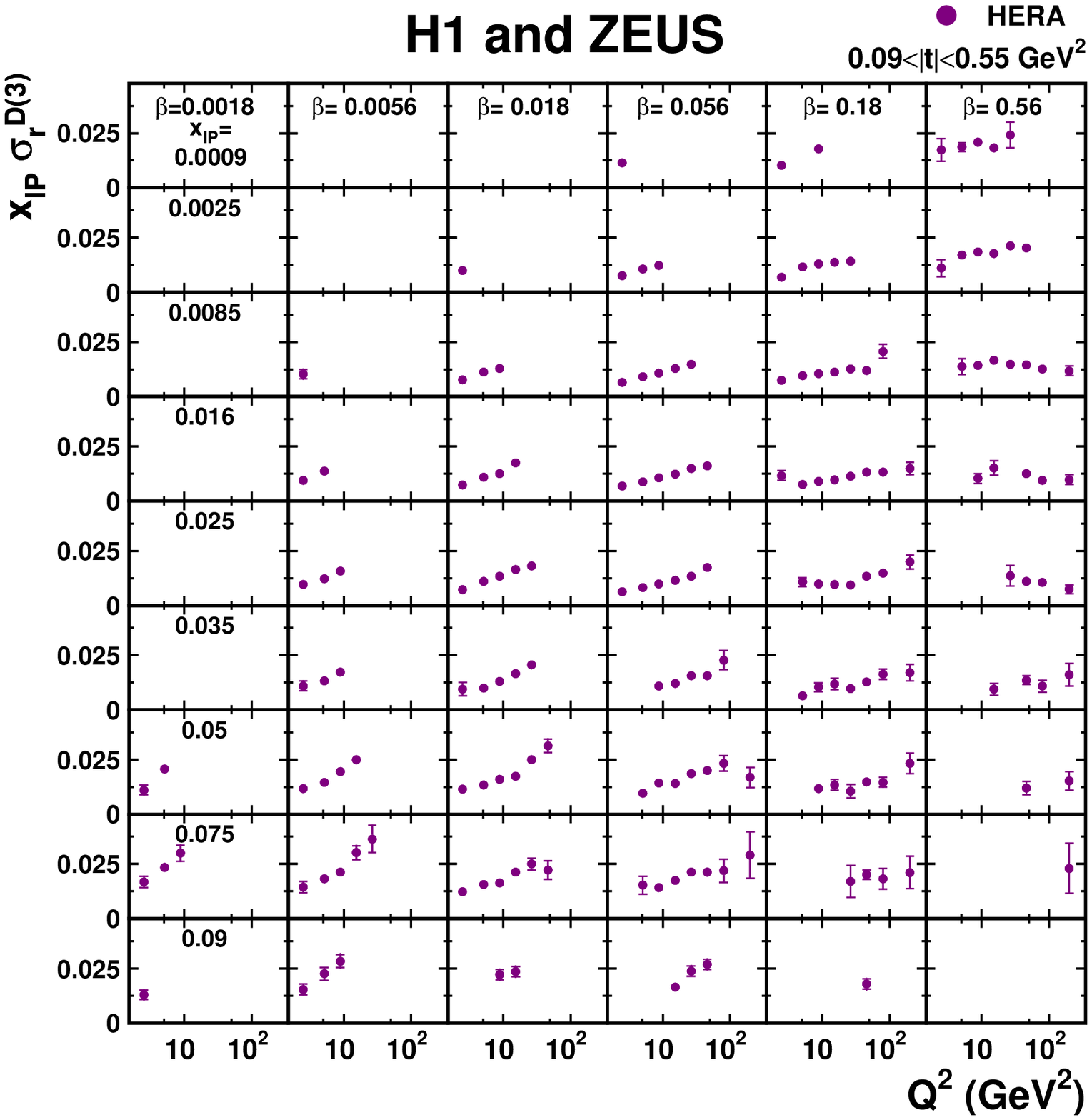}
\includegraphics[width=0.5\textwidth]{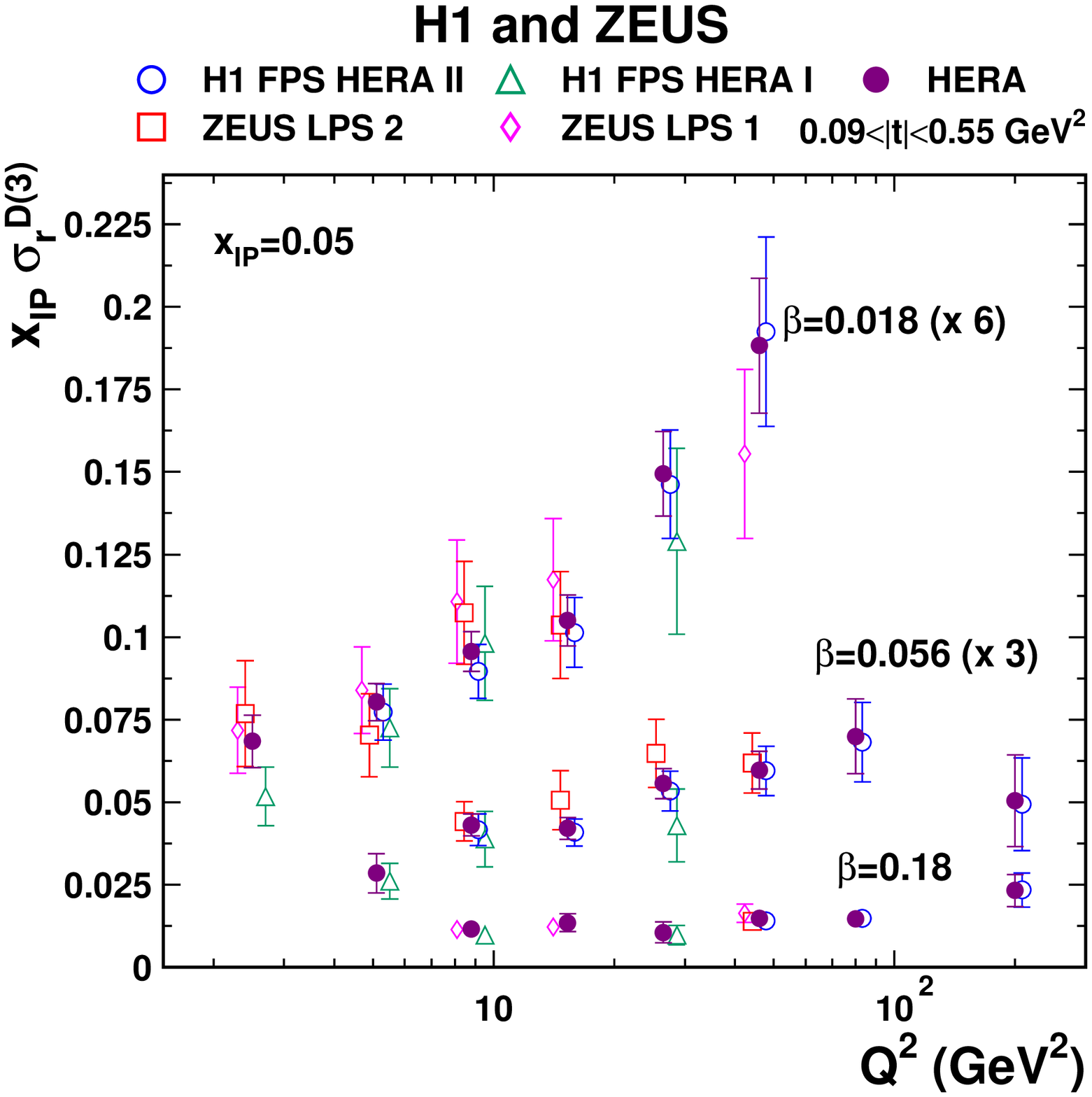}}
\caption{ HERA combined reduced diffractive cross section  $\xP\sigma_r^{D(3)}$ .}\label{FigureLabel}
\end{figure}

% ****************************************************************************
% BIBLIOGRAPHY AREA
% ****************************************************************************

\begin{footnotesize}
% IF YOU DO NOT USE BIBTEX, USE THE FOLLOWING SAMPLE SCHEME FOR THE REFERENCES
% ----------------------------------------------------------------------------

\end{footnotesize}
\end{document}